\documentclass{svproc}
\usepackage{url}
\usepackage{graphicx,bm,color}
\usepackage{slashed,verbatim}
\usepackage[english]{babel}
\usepackage{amsmath}
\usepackage{graphicx}
\usepackage[colorinlistoftodos]{todonotes}
\usepackage{amssymb,graphicx}

\def\be {\begin{equation}}
\def\ee {\end{equation}}
\def\bea {\begin{eqnarray}}
\def\eea {\end{eqnarray}}
\def\bc {\begin{center}}
	\def\ec {\end{center}}
\def\nn {\nonumber}
\def\eps {\epsilon}

\def\mn {\mu\nu}

\def\({\left(}
\def\){\right)}
\def\[{\left[}
\def\]
\def\sp{\shortparallel}

\def\sp {\shortparallel}

\def\sumintb{\sum\!\!\!\!\!\!\!\!\!\int\limits}
\def\sumintf{\sum\!\!\!\!\!\!\!\!\!\!\int\limits}

\def\slashed{\slash\!\!\!\!}

\begin{document}
\mainmatter              % start of a contribution
\title{Anisotropic pressure and quark number
susceptibility of strongly magnetized QCD medium}
\titlerunning{Anisotropic pressure and $\cdots$}  % abbreviated title (for running head)
\author{Bithika Karmakar\inst{1,4} \and Ritesh Ghosh\inst{1,4} \and
Aritra Bandyopadhyay\inst{2} \and Najmul Haque\inst{3,4} \and Munshi G Mustafa\inst{1,4}}
\authorrunning{Bithika Karmakar et al.} % abbreviated author list (for running head)
%Craig Chambers, Kim B. Bruce, and Elisa Bertino}
%
\institute{Theory Division, Saha Institute of Nuclear Physics, 
 	1/AF, Bidhannagar, Kolkata 700064, India,\\
\email{bithika.karmakar@saha.ac.in},\\ 
\and
Guangdong Provincial Key Laboratory of Nuclear Science, Institute of Quantum Matter,
South China Normal University, Guangzhou 510006, China,\\
\and
School of Physical Sciences, National Institute of Science Education and Research, Jatni, Khurda 752050, India,\\
\and
Homi Bhabha National Institute, Anushaktinagar,
Mumbai, Maharashtra 400094, India.}

\maketitle              % typeset the title of the contribution

\begin{abstract}
In this work, we compute the hard thermal loop pressure of quark-gluon plasma within strong magnetic field
approximation at one-loop order. Magnetic field breaks the rotational symmetry of the system. As a
result, the pressure of QGP becomes anisotropic and one finds two different pressures along the longitudinal
(along the magnetic field direction) and transverse direction. Similarly, the second-order quark number
susceptibility, which represents the fluctuation of the net quark number density, also becomes anisotropic.
We compute the second order QNS of deconfined QCD matter in strong field approximation considering same
chemical potential for two quark flavors.

\end{abstract}
\section{Introduction}
Recent studies have shown that the magnetic field created by the spectator particles at heavy ion collisions can be as high as $eB\sim 10^{18}$ Gauss~\cite{Skokov:2009qp,Kharzeev:2007jp} at the time of collision. The energy of quarks get quantized in the presence of strong magnetic field. In strong field approximation, only the lowest Landau level of quark energy is considered. The EoS of the system in presence of strong magnetic field is particularly important as it is used to find the time evolution of the sytem using various hydrodynamic models. Besides, quark number susceptibility represents the fluctuation of the net quark number density over the average value. Fluctuation of the conserved quantities are used as probe of the hot and dense matter created in heavy ion collisions. Here, we calculate the pressure and second order diagonal QNS of the strongly magnetized medium using one loop HTL pt. theory within the scale hierarchy $gT<T<\sqrt{|q_fB|}$.

\section{Formalism}
In thermal medium the presence of the heat bath velocity $u^\mu$ breaks the boost symmetry of the system. Moreover, the presence of magnetic field breaks the rotational symmetry of the system. We consider the rest frame of the heat bath velocity $u^\mu=(1,0,0,0)$ and the magnetic field along $z$ direction {\it{i.e.,}} $n_\mu=(0,0,0,1)$.       %Quarks are directly affected by the background magnetic field and the gluons are affected via the quark loop. 
% The parallel and perpendicular components of momentum in presence of magnetic field are given as
% \bea
% P^\mu_{\sp}=(P \cdot u)u^\mu-(P \cdot n)n^\mu\,,\,P^\mu_\perp=P^\mu-P^\mu_{\sp}.
% \eea
% The parallel and perpendicular components of the metric tensor are given as
% \bea
% \eta^{\mn}_{\sp}=u^\mu u^\nu-n^\mu n^\nu\,,\,\eta^{\mn}_\perp= \eta^{\mn}-\eta^{\mn}_{\sp}.
% \eea
\subsection{General structure of gluon effective propagator}
 Now, the gluon self energy $\Pi^{\mn}$ can be expressed in terms of the seven available tensors $P^\mu P^\nu, u^\mu u^\nu, n^\mu n^\nu, P^\mu u^\nu+u^\mu P^\nu, P^\mu n^\nu+n^\mu P^\nu, u^\mu n^\nu+n^\mu u^\nu$. Using the transversality condition $P^\mu \Pi_{\mn}=0$, the gluon self energy can be written~\cite{Karmakar:2018aig}
% \bea
% \Pi^{\mn}=\alpha B^{\mn}+\beta R^{\mn} +\gamma Q^{\mn} +\delta N^{\mn},
% \eea
in terms of the four constituent tensors $B^{\mn}, R^{\mn}, Q^{\mn}$ and $N^{\mn}$ which are given in Ref.~\cite{Karmakar:2018aig}.
% \bea
% B^{\mn}&=&\frac{\bar u^\mu \bar u^\nu}{\bar u^2},\\
% R^{\mn}&=&\eta_\perp^{\mn}-\frac{P_\perp^\mu P_\perp^\nu}{P_\perp^2},\\
% Q^{\mn}&=& \frac{\bar n^\mu \bar n^\nu}{\bar n^2},\\
% N^{\mn}&=& \frac{\bar u^\mu \bar n^\nu+\bar n^\mu \bar u^\nu}{\sqrt{\bar u^2}\sqrt{\bar n^2}},
% \eea
% with $\bar u^\mu=u^\mu-\frac{(P \cdot u)P^\mu}{P^2}$, $\bar n^\mu= n^\mu-\frac{(P \cdot n)P^\mu}{P^2}-\frac{( \bar u \cdot n)\bar u^\mu}{\bar u^2}$.
%

Using Dyson-Schwinger equation, the general structure of gluon effective propagator can be written as
\bea
\mathcal{D}_{\mn} &=&\frac{\xi P_{\mu}P_{\nu}}{P^4}+\frac{(P^2-\gamma) B_{\mn}}{(P^2-\alpha)(P^2-\gamma)-\delta^2}+\frac{R_{\mn}}{P^2-\beta}\nn\\
&&+\frac{(P^2-\alpha) Q_{\mn}}{(P^2-\alpha)(P^2-\gamma)-\delta^2}+\frac{\delta N_{\mn}}{(P^2-\alpha)(P^2-\gamma)-\delta^2}.
\eea
\subsection{General structure of quark effective propagator}
Similarly, the general structure of fermion self energy in the presence of strong magnetic field can be written as~\cite{Karmakar:2019tdp}
 \begin{align}
\Sigma(P) &= a \,\slashed{u} +b\,\slashed{n} + c\gamma_{5}\,\slashed{u} 
+d\gamma_{5}\,\slashed{n}. \nn
\end{align}
where the form factors in presence of strong magnetic field are calculated in Ref.~\cite{Karmakar:2019tdp}.
% \bea
% a&=&\frac{1}{4}\Tr[\Sigma \slashed{u}]\,,\,b=-\frac{1}{4}\Tr[\Sigma \slashed{n}],\\
% c&=&\frac{1}{4}\Tr[\gamma_5\Sigma \slashed{u}]\,,\,d=-\frac{1}{4}\Tr[\gamma_5\Sigma \slashed{n}].
% \eea
The general structure of fermion effective propagator is given as

 \bea
S_{eff}(P)&=& P_{-}\frac{\slashed{L}(P)}{L^2} P_{+}+P_{+}\frac{\slashed{R}(P)}{R^2} P_{-}
\eea
with
\bea
L^{\mu}&=&P^\mu +(a+c)u^{\mu}+(b+d)n^{\mu},\\
R^{\mu}&=&P^\mu +(a-c)u^{\mu}+(b-d)n^{\mu}.
\eea

\section{Anisotropic pressure and second order QNS of quark gluon plasma in strongly magnetized medium}

The pressure of quark gluon plasma in the presence of strong magnetic field becomes anisotropic {\it{i.e.,}} the pressure along the magnetic field (longitudinal) and the pressure perpendicular (transverse) to it becomes different.  In this case the free energy density in a  finite spatial volume $V$ is given by 
\bea
F=\mathcal{F}/V=\eps^{\text{total}}-Ts-eB\cdot M,
\eea
where $\eps^{\text{total}}$ ans $s$ are total the energy density and  entropy density.
% \bea
% s=-\frac{\partial F}{\partial T}\label{entropy}, 
% \eea
The magnetization per unit volume is given by
\bea
M=-\frac{\partial F}{\partial (eB)}. \label{magnetization}
\eea
Now, the longitudinal and transverse pressures are given as 
\bea
P_{z}=-F ,\,\,\, P_{\perp}=-F-eB\cdot M=P_z-eB\cdot M.\label{trans_pressure}
\eea

The free energy of quarks and gluons in the presence of strong magnetic field can be calculated using HTL approximation as

\bea
 F_q&=&-N_c N_f\sum_f \sumintf \,\,\ln {(\det[S_{eff}^{-1}])}\nn\\
 &=& -N_c N_f \sum_f \sumintf \,\,\ln \Big[{P_\sp^4 \Big(1+\frac{4a^2-4b^2+4ap_0+4bp_3}{P_\sp^2} \Big)}\Big].
\eea

%   where $S_{eff}^{-1}=\slash\!\!\!\! P +\Sigma$.
 
\bea
 F_g&=&(N_c^2-1) \bigg[\frac{1}{2}\sumintb \,\,\,\ln{[\det(D_{\mn}^{-1})]}-\sumintb \,\,\,\ln (-P^2)\bigg].\nn
\eea

Similarly, the second-order longitudinal and transverse QNS are defined as~\cite{Karmakar:2020mnj}
 \bea
\chi_z= \frac{\partial^2P_z}{\partial \mu^2}\bigg \vert_{\mu=0}\,,\,\chi_\perp= \frac{\partial^2P_\perp}{\partial \mu^2}\bigg \vert_{\mu=0}.
\eea

\section{Results}

\begin{center}
\begin{figure}[tbh!]
 \begin{center}
 \includegraphics[scale=0.29]{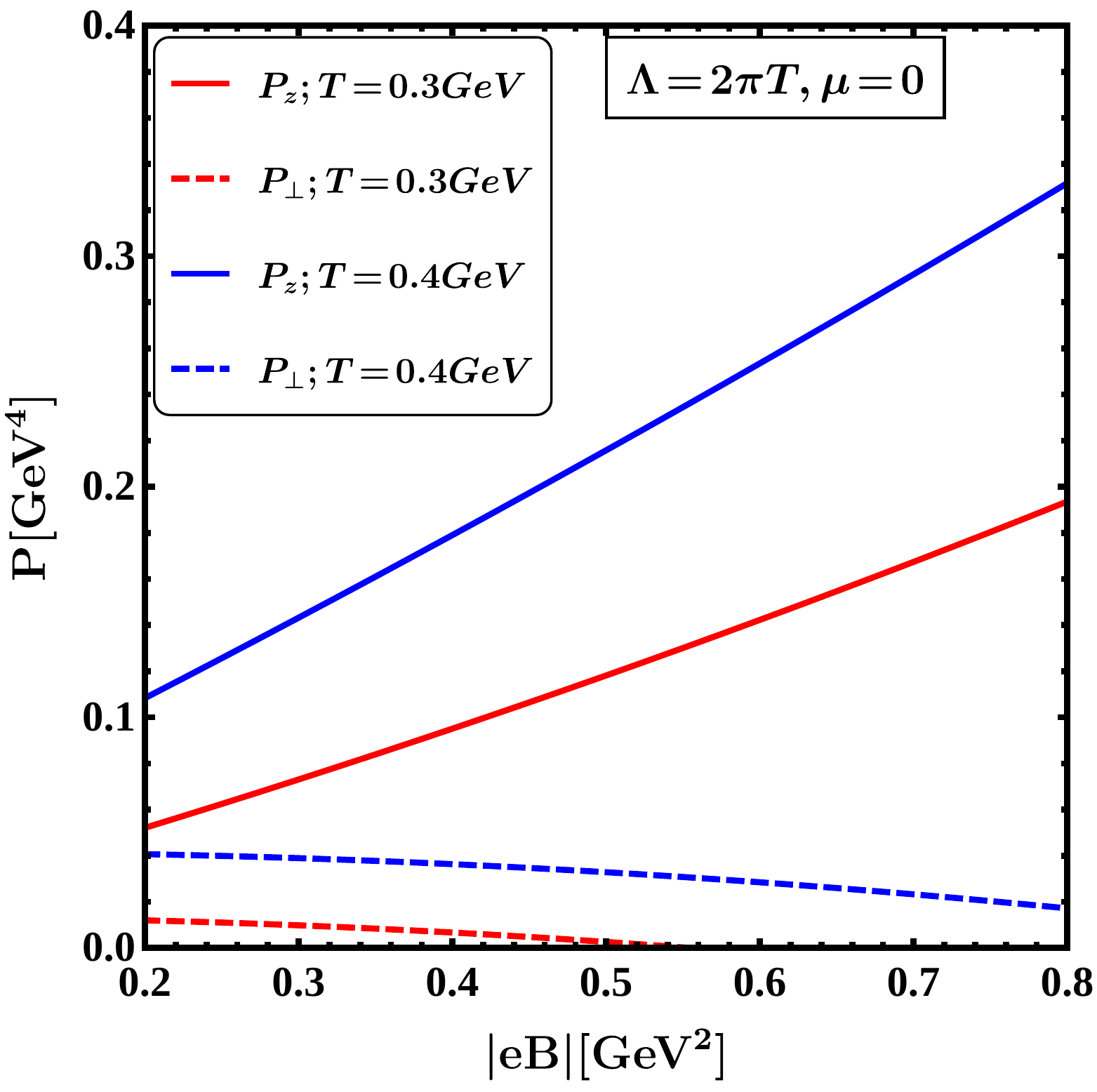} 
 \includegraphics[scale=0.36]{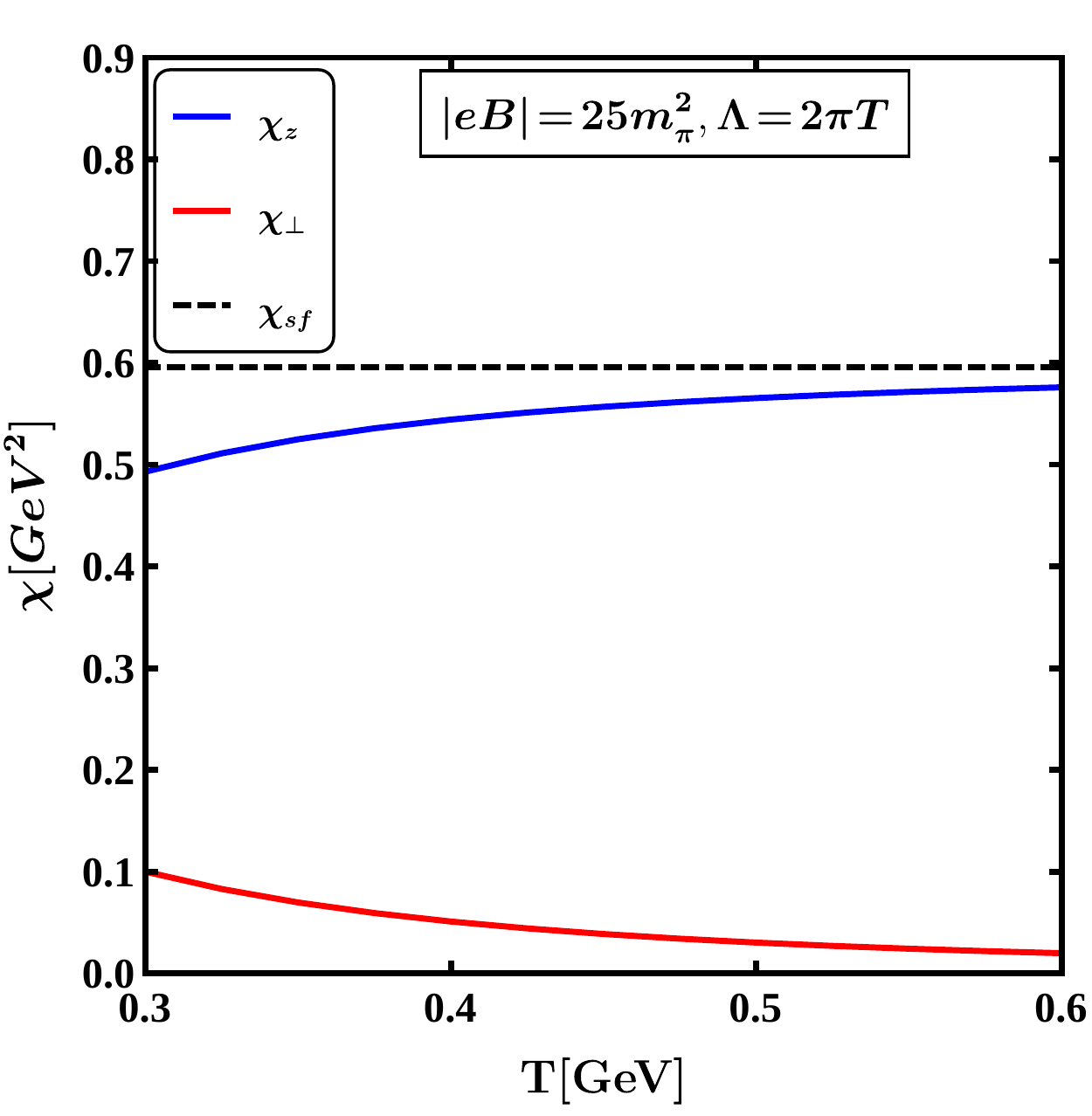} 
 \caption{Variation of the longitudinal and transverse pressure at $\mu=0$ with magnetic field is shown in left panel. Magnetization as a function of temperature at $\mu=0$ is shown in right panel for $N_f=3$.}
  \label{P_mag_sfa}
 \end{center}
\end{figure}
\end{center} 

In left panel of Fig.~\ref{P_mag_sfa}, the longitudinal and transverse pressure of strongly magnetized QGP are shown with the variation of magnetic field strength at $\mu=0$. It can be seen that the longitudinal and transverse pressure show different nature with magnetic field. The longitudinal pressure increases with the magnetic field whereas the transverse one decreases. This means that the system elongates more along the magnetic field direction and it can even shrink~\cite{Bali:2014kia} along the transverse direction at very high magnetic field strength. 

The variation of second order diagonal QNS of strongly magnetized QGP is shown with temperature in the right panel of Fig.~\ref{P_mag_sfa}. Similar to the anisotropic pressure, the longitudinal and the transverse QNS also shows different nature. At very high temperature, the QNS of the interacting system reaches the ideal value.

% 
% \begin{center}
% \begin{figure}[tbh!]
%  \begin{center}
%  \includegraphics[scale=0.35]{chi2_sfa_long_trans.pdf} 
%  \caption{Variation of the longitudinal and transverse QNS with temperature at magnetic field strength $25m_\pi^2$ is shown for $N_f=3$.}
%   \label{chi_sfa}
%  \end{center}
% \end{figure}
% \end{center} 
% 

%
\section{Summary}

Using one loop HTL pt theory with strong magnetic field approximation we found that the QGP pressure becomes anisotropic. Due to the presence of magnetization, the transverse pressure decreases with magnetic field. The transverse pressure can even be negative at very high magnetic field strength indicating the shink of the system along the transverse direction. It has been found that the system shows paramagnetic nature. On the other hand, second order diagonal QNS has been calculated using the longitudinal and the transverse pressure. We find qualitative match of our results with lattice studies~\cite{Bali:2014kia}. However, perturbative studies can be improved by considering higher loop order calculations.

\end{document}